\documentclass[10pt,twocolumn,letterpaper]{article}

\usepackage[pagenumbers]{wacv} 

\usepackage{graphicx}
\usepackage{amsmath}
\usepackage{amssymb}
\usepackage{booktabs}
\usepackage{adjustbox}
\usepackage{multirow}
\usepackage[accsupp]{axessibility}  

\usepackage{algorithm2e}
\usepackage{algpseudocode}
\RestyleAlgo{ruled}
\LinesNumbered
\SetKw{Continue}{continue}

\usepackage[pagebackref,breaklinks,colorlinks]{hyperref}

\usepackage[capitalize]{cleveref}
\crefname{section}{Sec.}{Secs.}
\Crefname{section}{Section}{Sections}
\Crefname{table}{Table}{Tables}
\crefname{table}{Table}{Tabs.}

\def\wacvPaperID{624} 
\def\confName{WACV}
\def\confYear{2025}

\usepackage{xcolor}
\definecolor{rebuttalcolor}{RGB}{0,165,135}

\newcommand*{\method}[1]{NCAdapt} 

\begin{document}

\title{\method{}: Dynamic adaptation with domain-specific Neural Cellular Automata for continual hippocampus segmentation}

\author{Amin Ranem$^{1}$\\
\and 
John Kalkhof$^{1}$\\
\and
Anirban Mukhopadhyay$^{1}$\\
\and
{\small$^{1}$ Technical University of Darmstadt}, 
{\tt\small amin.ranem@gris.tu-darmstadt.de}}

\maketitle

\begin{abstract}
Continual learning (CL) in medical imaging presents a unique challenge, requiring models to adapt to new domains while retaining previously acquired knowledge. We introduce \textbf{\method{}}, a Neural Cellular Automata (NCA) based method designed to address this challenge. \method{} features a domain-specific multi-head structure, integrating adaptable convolutional layers into the NCA backbone for each new domain encountered. After initial training, the NCA backbone is frozen, and only the newly added adaptable convolutional layers, consisting of 384 parameters, are trained along with domain-specific NCA convolutions. We evaluate \method{} on hippocampus segmentation tasks, benchmarking its performance against Lifelong nnU-Net and U-Net models with state-of-the-art (SOTA) CL methods. Our lightweight approach achieves SOTA performance, underscoring its effectiveness in addressing CL challenges in medical imaging. Upon acceptance, we will make our code base publicly accessible to support reproducibility and foster further advancements in medical CL.
\end{abstract}

\section{Introduction}
In medical continual learning (CL), U-Net architectures have long been favored for their proven effectiveness \cite{chen2021transunet,pmlr-v172-fuchs22a,gonzalez2023lifelong,isensee2021nnu,ranem2022continual}. However, their static nature presents significant challenges in adapting to the evolving dynamics of medical data. Models with ever-increasing parameter counts are conventionally employed to increase performance in static settings. Yet they struggle to adapt effectively to shifts in data distribution, limiting their applicability in dynamic medical imaging scenarios.

Efforts such as the nnU-Net aim to introduce self-adjustment capabilities but encounter difficulties in CL settings \cite{isensee2021nnu,gonzalez2022task,ranem2022continual,ranem2024continual}. Extensions like Lifelong nnU-Net have sought to enhance adaptability using techniques such as Elastic Weight Consolidation (EWC, \cite{kirkpatrick2017overcoming}) and Riemannian Walk (RWalk, \cite{chaudhry2018riemannian}). However, their reliance on over-parametrization can slow down adaptation processes \cite{gonzalez2020wrong,ranem2022continual}. Persistent challenges remain in accommodating shifts in data distribution, especially with new imaging protocols and variations in diseases and input sizes \cite{derakhshani2022lifelonger,pmlr-v172-fuchs22a,gonzalez2022task,sanner2021reliable}. In medical continual learning (CL), strategies like replay, regularization, or knowledge distillation are promising but often impractical. For example, rehearsal methods, which involve storing and revisiting previous data, are typically infeasible due to privacy concerns and the massive data storage requirements in medical contexts. These strategies also introduce complexities such as balancing rigidity and adaptability or increasing computational demands \cite{gonzalez2020wrong}. Therefore, there is a need for a dynamic model capable of learning new tasks while retaining past knowledge \cite{de2021continual,hadsell2020embracing,kirkpatrick2017overcoming}.

Recognizing the critical need for adaptability in CL frameworks, we advocate a departure from U-Net architectures towards Neural Cellular Automata (NCA) \cite{gilpin2019cellular,mordvintsev2020growing}. NCAs offer a promising alternative by operating on a set of rules uniformly applied across all cells, facilitating dynamic communication among neighboring cells while maintaining a significantly smaller model size \cite{kalkhof2023med,kalkhof2023m3d}. This adaptability makes NCAs particularly suited for handling sequential and evolving data. Their resilience in dynamic environments allows them to effectively address the challenges encountered in adapting to diverse medical imaging scenarios.

\begin{figure*}[htp]
    \centering
    \includegraphics[trim=5cm 6cm 6cm 6cm, clip, width=\textwidth]{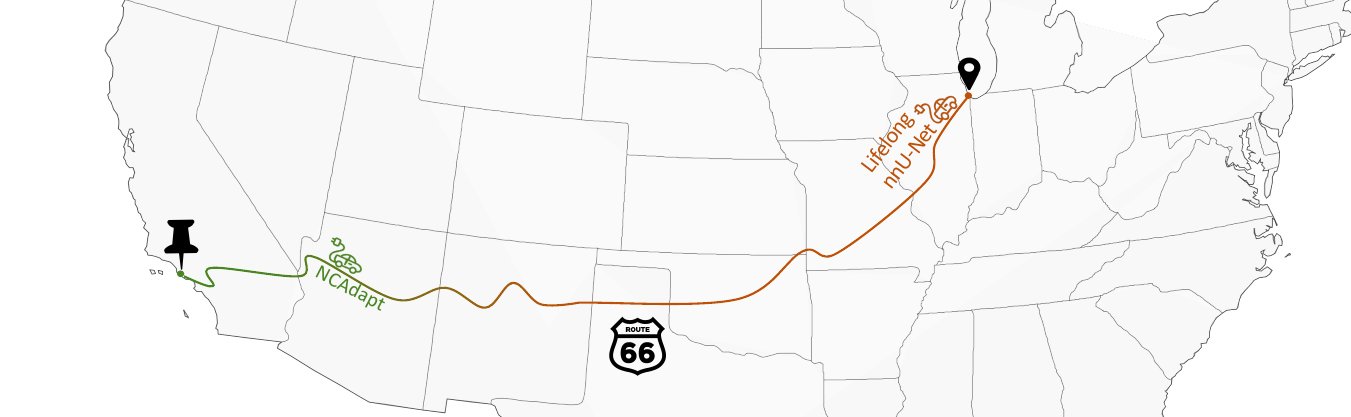}
    \caption{Simulation of 467 training stages comparing the estimated carbon footprint between Lifelong nnU-Net and \method{}. To estimate per stage, the codecarbon python package is used \cite{codecarbon} (Lifelong nnU-Net: 616 kWh (entire Route 66 of 3940 km); \method{}: 122 kWh). The route is based on the average range of a Tesla Model 3, where 1 kWh equals 6.4 km (4 miles).}
    \label{fig:intro}
\end{figure*}

We introduce \textbf{\method{}}, a robust and energy-efficient approach harnessing NCAs for continuous setups in medical image segmentation. Unlike traditional large-scale models like nnU-Nets, \method{} emphasizes task-specific layers with a minimal parameter count of only 6,339 trainable parameters. Figure \ref{fig:intro} illustrates the significantly lower energy consumption of our lightweight \method{} compared to Lifelong nnU-Net across 467 continuous training stages. This design enables ongoing adaptation of domain-specific convolutional layers within the NCA backbone, mitigating issues associated with catastrophic forgetting.

\method{} enhances adaptability to shifting domains, ensuring efficient adaptation without necessitating full model re-training or complex regularization techniques. By integrating domain-specific layers to acquire new domain knowledge while freezing the NCA backbone after initial domain training, \method{} strikes a delicate balance between rigidity and adaptability, effectively addressing challenges in domain adaptation posed by contemporary U-Net-based architectures. Our contributions are threefold: \method{} (1) is \textbf{energy-efficient, while using significantly less parameters than (nn)U-Nets} and (2) \textbf{adaptable to various image sizes} for continual medical image segmentation. \method{} (3) excels in \textbf{adapting to shifting domains, achieving state-of-the-art (SOTA) performance}, swiftly training on new domains, and demonstrating superior applicability. To validate our approach, we focus on continual hippocampus segmentation for T1-weighted Magnetic Resonance Images (MRIs), crucial for diagnosing and selecting optimal treatments for neuropsychiatric disorders susceptible to distribution shifts.

\section{Fundamentals on NCAs}
Neural Cellular Automata (NCA) \cite{mordvintsev2020growing} represent a departure from traditional U-Net-based deep learning models by embodying a learned local update rule. This rule governs how each cell's state evolves based on its current state and the states of its neighboring cells. Inspired by biological processes like morphogenesis, NCAs exhibit emergent behavior that can be dynamically learned over iterations across space and time, making them well-suited for complex tasks such as medical image segmentation \cite{kalkhof2023med}. This adaptability mirrors biological systems, which naturally adjust to new environmental conditions; a crucial advantage in medical domains where data can vary significantly due to different imaging protocols, disease manifestations, and patient demographics.

\paragraph{NCA specific concepts}
Due to their vastly different approach, there are different kinds of concepts to be considered. The number of steps, defines the \emph{perceptive range}, as an NCA can only communicate with direct neighbors, it can only communicate one cell at a time, requiring the iterative application of the update rule to gather global information. To store knowledge in between steps, each cell has a state $s$, which consists of typically 16 values for storing information. NCAs generally show great robustness, which is linked to the stochastic activation of each cell, defined by the fire rate which typically sits at 50\%. Due to this fire rate, the model has to learn to deal with very different update paths, fostering a robust learned rule.

\paragraph{NCA Quality Metric (NQM)}
Stochastic activation in NCAs allows the model to produce slightly different predictions for the same input each time it runs, due to varying inference paths. This variability means that models with less robust or accurate predictions will show greater differences between these outputs. We can leverage this variance to estimate the quality of the model’s predictions.\\
To quantify the quality of a segmentation mask, we use the NQM score \cite{kalkhof2023m3d}. The NQM score improves prediction by normalizing the variance based on the size of the segmentation mask. For each input image $x_i$, we generate $N$ predictions from each domain-specific head. We then calculate the standard deviation (SD) $SD = \sqrt{\frac{\sum_{i = 1}^{N}(h_j(x_i) - \mu)^2}{N}}$ and mean ($\mu$) of these predictions $\mu = \frac{\sum_{i = 1}^{N}h_j(x_i)}{N}$. The NQM score is then computed as:
\begin{align}
\label{eqn:NQM}
NQM &= \frac{\sum_{s \in SD}(s)}{\sum_{m \in \mu}(m)}.
\end{align}

\begin{figure*}[ht!]
    \centering
    \includegraphics[width=\textwidth]{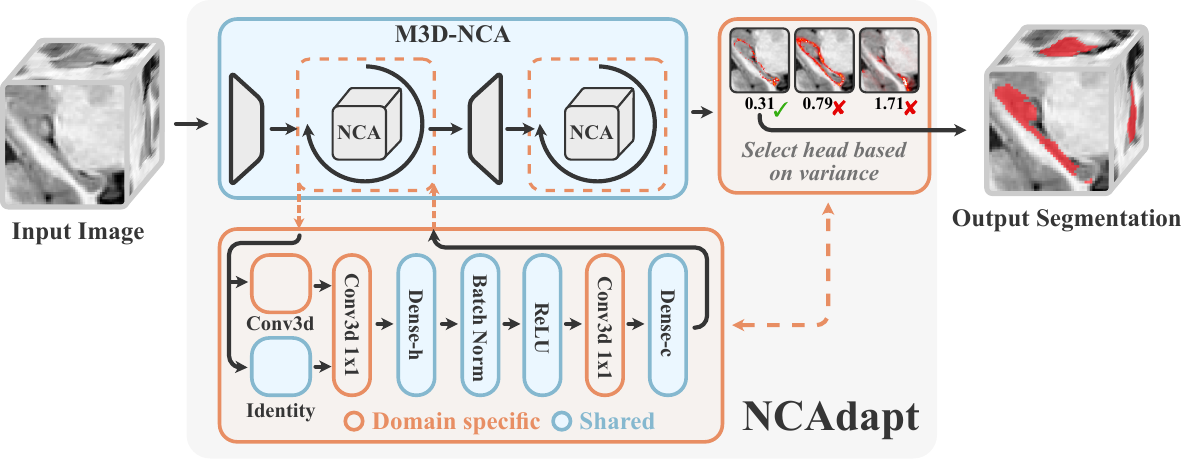}
    \caption{\method{} for medical image segmentation using a M3D-NCA with an M3D-NCA as a backbone while having shared layers and domain-specific convolutional layers.}
    \label{fig:method}
\end{figure*}

This score measures the consistency and reliability of the segmentation predictions by analyzing the variance across multiple predictions. It provides valuable insights into the robustness and accuracy of the model's outputs.

\section{Methodology}
We propose \method{}, a continual NCA-based method for medical image segmentation that is adaptable to domain shifts over time. 

\paragraph{M3D-NCA} \cite{kalkhof2023m3d} employs an n-level approach where multiple NCAs operate on different scales of the input image, spanning from coarse to fine perspectives. This design facilitates global information exchange being faster in the early levels. The global information is then upsampled to subsequent image scales, enabling integration with high-resolution image details. \textit{\method{} utilizes the M3D-NCA \cite{kalkhof2023m3d} model as its backbone}.

\paragraph{Training continuously with \method{}} Unlike traditional approaches, \method{} adapts itself by incorporating domain-specific convolutional layers, which are crucial to leverage and maintain domain-knowledge. \method{} is trained for the first stage to acquire general information on the segmentation task, after which the NCA backbone is frozen. \method{} only uses 6,336 trainable parameters and is \textit{2,559-times smaller than the nnU-Net}. This frozen state \textit{ensures stability and consistency in feature extraction}, providing a reliable baseline for subsequent adaptation. Concurrently, \textit{domain-specific convolutional layers are continuously trained} and updated to capture new domain knowledge and therefore \textit{preventing catastrophic forgetting} over time. Using this method, \method{} increases the number of parameters by \textbf{only 3\% (384)} with the detection of each new domain, while the amount of trainable parameters remains the same (6,336 in total). These layers dynamically adjust their parameters based on the input data, allowing the model to adapt seamlessly to changing domain characteristics \textit{without the need for retraining from scratch for each new domain}, Figure \ref{fig:method}.

The domain-specific layers within our NCA approach play a crucial role in learning and capturing new domain knowledge. These layers are specifically tailored to extract and encode domain-specific features, enabling the model to effectively reuse the acquired knowledge during inference, while minimizing the amount of forgetting. By focusing on learning domain-specific information \textit{with a frozen NCA backbone, \method{} is light-weight, adaptable and robust} across different datasets. This adaptability ensures that the model can effectively handle variations in imaging protocols, patient demographics, input sizes and disease characteristics, thereby maintaining high segmentation performance in dynamic clinical environments.

\paragraph{Inference using \method{}} During inference, when task-specific information is unavailable, we use a refined method that incorporates all domain-specific heads along with the NCA Quality Metric (NQM) \cite{kalkhof2023m3d}. This approach is variance-based and normalized for segmentation mask size, enabling the selection of the optimal prediction (see Equation \ref{eqn:NQM}).

We select the prediction $\hat{h}(x_i) = h_k(x_i)$, where $k = \arg \max_{j} \left( NQM_j(x_i) \right)$, prioritizing predictions that not only show minimal variance but also adjust for the segmentation mask size. This concise approach enhances accuracy by integrating a variance-based metric that is normalized, improving the reliability of model outputs.

This simple approach, ensures that the final prediction is robust and reliable based on the acquired knowledge from different domains. This enables \method{} to effectively adapt to diverse datasets and domain shifts while maintaining high prediction accuracy during inference.

In summary, \method{} represents a novel and effective solution for continual hippocampus segmentation. By leveraging the inherent capabilities of NCAs, combined with domain-specific convolutional layers and continuous adaptation, our method achieves superior adaptability and segmentation performance across evolving datasets and domain shifts. \method{} can train and predict on variable image dimensions which can not be done with default U-Net-based architectures.

\paragraph{Training continually using \method{}} Algorithm \ref{alg:train} demonstrates a continous training setup in simplified pseudo-code for our proposed NCAdapt approach using NCAs.

\begin{algorithm}[htp]
\caption{Training using \method{}}
\label{alg:train}
\KwIn{Datasets to train on $\{\mathcal{T}_{1}, \dots, \mathcal{T}_{n}\}$}
\KwOut{Trained model weights $\theta$}
\tcp{Initialize $\mathcal{M}_\theta$}
$\theta \gets initializeModel()$ 

\tcp{Train with NCA}
$\theta \gets train(\text{NCA}, \theta, \mathcal{T}_1)$

\tcp{Freeze NCA layers}
$\text{NCA} \gets freezeNonDomainLayers(NCA)$
    
\For{$i\leftarrow 2$ \KwTo $n$}{

    \tcp{Add new domain layers for $\mathcal{T}_i$}
    $\text{NCA} \gets addDomainLayers(NCA)$
    
    \tcp{Train with NCA}
    $\theta \gets train(\text{NCA}, \theta, \mathcal{T}_i)$
}
\end{algorithm}

\paragraph{Energy Efficiency} The design of \method{} ensures it is energy efficient, consuming significantly less power than nnU-Net models during continuous training. As depicted in Figure \ref{fig:intro}, the energy consumption for \method{} was mapped along the Route 66 simulation for 467 continuous training stages. This efficient energy consumption further supports its practicality in real-world medical applications where resources are often limited \cite{jia2023importance}.

\paragraph{Handling Diverse Image Dimensions} Another advantage of \method{} is its ability to handle variable image dimensions seamlessly. Traditional U-Net-based architectures often struggle with images of differing sizes, requiring resizing that can lead to loss of crucial information. \method{}'s flexible structure allows it to process images of various dimensions without compromising on the quality of segmentation, making it particularly useful for clinical settings where imaging protocols and equipment may vary over time.

By integrating these additional aspects, \method{} stands out as a comprehensive solution for CL in medical image segmentation, combining efficiency, adaptability, and robustness to meet the demands of dynamic and diverse clinical environments.

\section{Experimental Setup}

\paragraph{Datasets}
We explore the problem of continual hippocampus segmentation for brain and left ventricle segmentation for cardiac MRIs. To ensure reproducibility, we utilize openly accessible datasets, with each data source serving as an individual task $\{\mathcal{T}_1, \dots, \mathcal{T}_{n}\}$. Table \ref{tab:data} summarizes the core characteristics of the hippocampus datasets. Details regarding the used cardiac datasets \cite{campello2021multi} is provided in Appendix, Section 1.\\
The hippocampus data corpus consists of three publicly available T1-weighted MRI datasets with senior healthy subjects, patients with Alzheimer’s disease and schizophrenia patients; Harmonized Hippocampal Protocol data \cite{boccardi2015training} (HarP), Dryad \cite{kulaga2015multi} and DecathHip from the Medical Segmentation Decathlon \cite{antonelli2021medical}, see Table \ref{tab:data}. For all datasets, we randomly use $20\%$ from the data for test purposes and maintain this split across all experiments.

\begin{table}[htp]
\centering
\caption{Image and label characteristics of the used hippocampus datasets.}
\label{tab:data}
\begin{adjustbox}{width=\linewidth}{
\begin{tabular}{ccccccc}
\hline
\multicolumn{1}{l}{Dataset} & & HarP           & &  Dryad              & & DecathHip \\ \hline \hline
\multicolumn{1}{l|}{\# Cases}  & & 270              & & 50                & & 260   \\
\multicolumn{1}{l|}{Resolution} & & {[}48 64 64{]}    & & {[}48 64 64{]}    & & {[}36 50 35{]}  \\
\bottomrule
\end{tabular}
}
\end{adjustbox}
\end{table}

\paragraph{Training setup}
All Lifelong nnU-Net \cite{isensee2021nnu,gonzalez2023lifelong} experiments are trained for 250 epochs using the framework's default setup. For the U-Net and M3D-NCA related experiments, we train for 500 epochs following the default setup and loss function of M3D-NCA \cite{kalkhof2023m3d}. The Adam optimizer with betas of $(0.9, 0.99)$, a learning rate of $16e^{-4}$, and an exponential learning rate scheduler with a gamma of $0.9999$ is used. Dice Focal Loss is the loss function for M3D-NCA, and Dice Binary Cross Entropy loss is used for U-Nets. For U-Nets, all training and testing images have been resampled to the input size of the first training task, as the U-Net architecture is not adaptive to varying image sizes. Each model trains individually for every task in the dataset to assess generalizability across tasks, providing insights for CL setups. All models are trained on a single GeForce RTX 3090 GPU (24 GB).

\begin{figure*}[ht!]
    \centering
    \includegraphics[width=\textwidth]{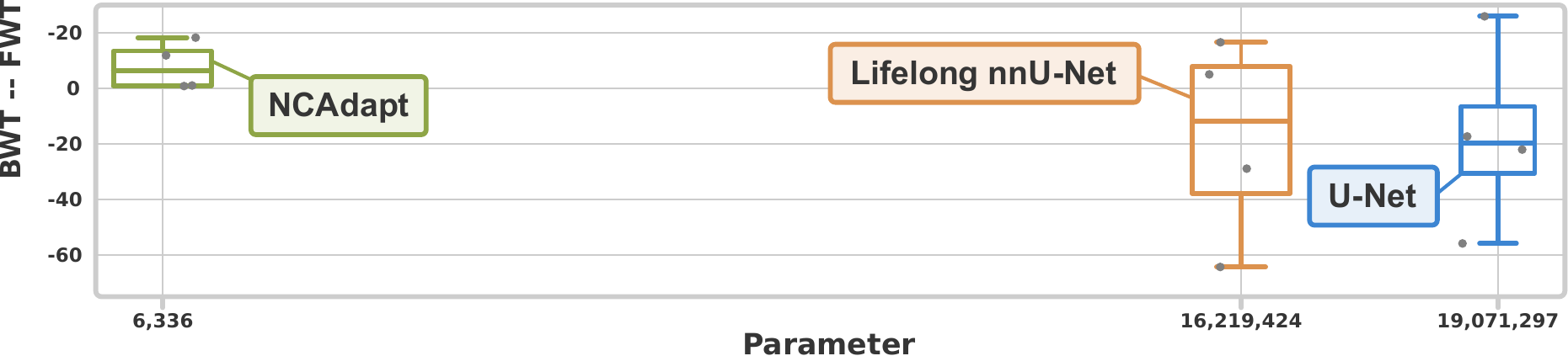}
    \caption{Comparison of BWT and FWT performance against number of parameters for \method{}, sequential nnU-net and U-Net; smaller boxes indicate superior performance. Note: Values above zero on the y-axis represents FWT and values below BWT.}
    \label{fig:bwt_fwt}
\end{figure*}

\paragraph{Domain Incremental Learning (DIL)} in medical image segmentation involves adapting machine learning models to continually learn and incorporate new domain knowledge over time \cite{joshi2012incremental}. This is crucial to handle shifts in data acquisition protocols and imaging technologies over time \cite{gonzalez2020wrong}. Unlike traditional static models that are trained on fixed datasets, DIL methods dynamically update their knowledge to accommodate variations in imaging protocols, disease presentations, and patient demographics \cite{ranem2024continual}. This adaptability is crucial in clinical settings where variations in imaging protocols, disease manifestations, and patient demographics necessitate models capable of flexibly updating their knowledge without forgetting previously learned information \cite{gonzalez2023lifelong}. Key methodologies in DIL include strategies for managing domain shifts, CL frameworks that prioritize robustness to new data distributions, and techniques for preserving model performance across evolving medical imaging scenarios.

\paragraph{Metrics}
For every CL setup, we report average backwards (BWT) and forwards (FWT) transferability \cite{diaz2018don} based on the mean Dice and standard deviation across the test images from all tasks as done in the Lifelong nnU-Net framework \cite{gonzalez2023lifelong}.
Let $\mathcal{T}_{i}$ be a specific task; 

FWT measures the impact of the current training stage $\mathcal{T}_i$ on test data from an untrained stage $\mathcal{T}_j\; ; \;j > i$ and is defined as: 

\begin{align}
\label{eqn:F}
    \text{FWT}\left( \mathcal{T}_{i}\right) &= \text{Dice}\left(\mathcal{M}_{\left[ \mathcal{T}_{1}, \dots, \mathcal{T}_{i-1}\right]}, \mathcal{T}_{i}\right) - \text{Dice}\left(\mathcal{M}_{\left[\mathcal{T}_{i}\right]}, \mathcal{T}_{i}\right),
\end{align}

where $\mathcal{M}_{\left[ \mathcal{T}_{1}, \dots, \mathcal{T}_{i}\right]}$ is a network trained on stages $\{1, \dots, p\}$ and $\text{Dice}(\mathcal{M}_{\left[ \mathcal{T}_{1}, \dots, \mathcal{T}_{j}\right]}, \mathcal{T}_{i})$ indicates the Dice score from a network trained on stages $\{1, \dots, j\}$ evaluated on dataset $p$.

BWT on the other hand representes the amount of maintained knowledge on test samples from $\mathcal{T}_j$ during training on different stages $\{\mathcal{T}_i\}
\; ; \;j < i$ over time and is defined as
\begin{align}
\label{eqn:B}
    \text{BWT}\left( \mathcal{T}_{i}\right) &=
    \text{Dice}\left(\mathcal{M}_{\left[ \mathcal{T}_{1}, \dots, \mathcal{T}_{i}, \dots, \mathcal{T}_{n}\right]}, \mathcal{T}_{i}\right) \nonumber \\ &- \text{Dice}\left(\mathcal{M}_{\left[ \mathcal{T}_{1}, \dots, \mathcal{T}_{i}\right]}, \mathcal{T}_{i}\right).
\end{align}

Models with high BWT scores maintain most knowledge from previous tasks, i.e. prevent catastrophic forgetting, while models with a higher plasiticity achieve higher FWT results. 

FWT for the last model state as well as BWT for the first model state is not defined.

\paragraph{CL Baselines}
To get a proper evaluation of our approach, we compare against conventional sequential training for Lifelong nnU-Net with two well-known CL methods: EWC \cite{kirkpatrick2017overcoming} and RWalk \cite{chaudhry2018riemannian}. For both CL methods we use the standard Lifelong nnU-Net \cite{gonzalez2023lifelong} hyperparameter setup for all our experiments (EWC: $\lambda = 0.4$, RWalk: $\alpha = 0.9, \lambda = 0.4$). Additionally, we use four rehearsal-based CL methods along with a default 3D U-Net \cite{Pérez-García_2020}: Synaptic Intelligence (SI) \cite{zenke2017continual}, Function Distance Regularization (FDR) \cite{benjamin2018measuring}, Dark Experience Replay (DER) \cite{buzzega2020dark} and Averaged GEM (AGem) \cite{chaudhry2018efficient}. However, \textit{it is important to note that rehearsal-based approaches, which involve storing and replaying past data, present practical challenges in medical CL settings} \cite{ranem2022continual, gonzalez2023lifelong}. These challenges stem from the core principles of medical CL, which emerged under the assumption of limited access to early data and privacy concerns in medical contexts. As such, there is a critical need for CL methods that can effectively adapt to new domains and data distributions without relying on access to previous data domains.

\begin{figure*}[h!]
    \centering
    \includegraphics[clip, trim=0 4.5cm 0 0, width=\textwidth]{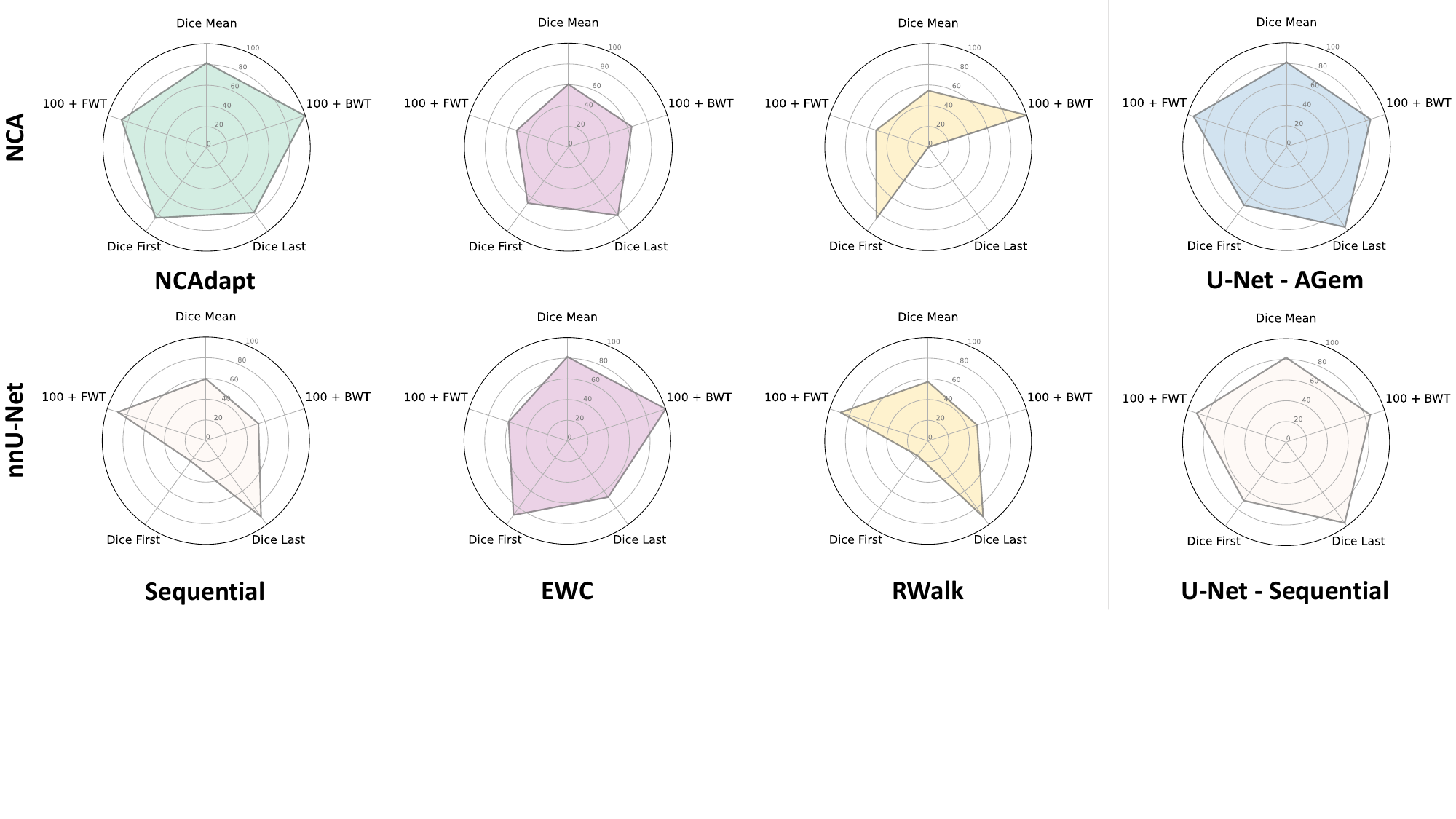}
    \caption{CL performance for \method{}, NCA with EWC and RWalk, Lifelong nnU-Net (sequential, EWC and RWalk) and U-Net (Sequential and AGem) using Dice scores with positive BWT and FWT; the larger the covered area the better the method.}
    \label{fig:spiders}
\end{figure*}
\begin{table*}[htb!]
\centering
\caption{CL performance of the final model; mean Dice, BWT and FWT over all tasks including standard deviation, total amount of trainable parameters, training runtime and inference time in seconds; best values for sequential setup are marked in bold. Methods marked with * are rehearsal-based methods. The inference time for nnU-Nets cannot be calculated as the input cases are processed in parallel.}
\label{tab:ps}
\begin{adjustbox}{max width=\linewidth}
{\begin{tabular}{lccccc|ccc}
\toprule
Method & Fixed param & Tuned param & Dice $\uparrow$ [\%] & BWT $\uparrow$ [\%] & FWT $\uparrow$ [\%] & \# Parameters (train) $\downarrow$ & Runtime [sec] $\downarrow$ & GPU sec $\downarrow$ \\ \midrule \midrule
$\text{Sequential}_{\text{nnU-Net}}$ & \multirow{4}{*}{--} & \multirow{4}{*}{--} & $59.56 \pm 27.15$ & $-46.61 \pm 17.70$ & $-10.80 \pm 5.78$ & 16,219,424 & $45.50$ & $--$ \\
$\text{Sequential}_{\text{U-Net}}$ & & & $\mathbf{81.75 \pm 11.05}$ & $-14.41 \pm 2.59$ & $\mathbf{-9.23 \pm 44.78}$ & 19,071,297 & $\mathbf{9.15}$ & $0.5$ \\
$\text{Sequential}_{\text{NCA}}$ & & & $29.41 \pm 40.88$ & $-86.38 \pm 3.55$ & $-45.09 \pm 38.15$ & 12,480 & $45.25$ & $0.55$\\
$\text{Sequential}_{\text{\method{}}}$ & & & $81.61 \pm 2.82$ & $\mathbf{-0.43 \pm 0.10}$ & $-13.70 \pm 3.20$ & \textbf{6,336} & $18.85$ & $\mathbf{0.49}$\\
\midrule
$\text{EWC}_{\text{nnU-Net}}$ & \multirow{2}{*}{--} & \multirow{2}{*}{$\lambda = 0.4$} & $81.36 \pm 9.97$ & $-46.77 \pm 12.16$ & $-52.72 \pm 16.90$ & 16,219,424 & $42.04$ & $--$ \\
$\text{EWC}_{\text{NCA}}$ & & & $60.55 \pm 19.61$ & $0.01 \pm 0.16$ & $-40.27 \pm 35.24$ & 12,480 & $39.80$ & $0.38$ \\
\midrule
$\text{RWalk}_{\text{nnU-Net}}$ & \multirow{2}{*}{\shortstack{$\alpha = 0.9$}} & \multirow{2}{*}{$\lambda = 0.4$} & $57.14 \pm 30.14$ & $-48.62 \pm 13.42$ & $-48.73 \pm 9.52$ & 16,219,424 & $75.67$ & $--$\\
$\text{RWalk}_{\text{NCA}}$ & & & $54.60 \pm 38.66$ & $-50.21 \pm 20.95$ & $-11.02 \pm 6.02$ & 12,480 & $37.22$ & $0.75$ \\
\midrule
$\text{SI}^{*}_{\text{U-Net}}$ & $c = 0.4$ & \multirow{4}{*}{--} & $78.25 \pm 13.77$ & $-19.10 \pm 4.52$ & $-18.56 \pm 52.44$ & \multirow{4}{*}{19,071,297} & $8.75$ & $0.45$ \\
$\text{FDR}^{*}_{\text{U-Net}}$ & -- & & $82.25 \pm 10.68$ & $-13.71 \pm 2.88$ & $-9.95 \pm 45.51$ & & $10.66$ & $0.65$\\
$\text{DER}^{*}_{\text{U-Net}}$ & $\alpha = 0.4$ & & $83.58 \pm 10.14$ & $-11.67 \pm 3.65$ & $-4.35 \pm 39.22$ & & $8.26$ & $0.47$\\
$\text{AGem}^{*}_{\text{U-Net}}$ & -- & & $81.27 \pm 10.84$ & $-14.88 \pm 2.06$ & $-5.65 \pm 40.77$ & & $7.90$ & $0.39$ \\
\bottomrule
\end{tabular}}
\end{adjustbox}
\end{table*}

\section{Results}
In Section \ref{sec:param}, we explore the CL performance against parameter efficiency of \method{}, highlighting its ability to achieve competitive results with fewer parameters. We comprehensively evaluate \method{} against SOTA CL approaches in Section \ref{sec:cl_res}, focusing on its adaptability and performance over time. Additionally, we analyze \method{}'s performance in static model scenarios in Section \ref{sec:base_models}, contrasting it with traditional methods like nnU-Nets. In Section \ref{sec:ablation} we perform an ablation study to dissect \method{}'s components and configurations, shedding light on critical design choices impacting its performance in continous settings. Lastly, in Section \ref{sec:qual}, we conduct a qualitative temporal evaluation, focusing on \method{}'s consistency and reliability over extended periods, crucial for sustained performance in practical applications. All results from the cardiac experiments are included in the supplementary material.

\subsection{CL performance and parameter efficiency}
\label{sec:param}

Figure \ref{fig:bwt_fwt} compares the CL performances to the number of trainable parameters across \method{}, the sequential Lifelong nnU-Net and U-Net. The CL performance is based on the positive FWT indicating values above the zero line and BWT reflecting negative values by default, i.e. values below the zero line. The performances deteriorate as the boxplots increase in size, indicating significant deviation in terms of CL performance. \method{} uses significantly fewer parameters compared compared to Lifelong nnU-Net and U-Net, while showing minimal deviation in FWT and BWT values, indicating a robust architecture over time.

\subsection{CL performance}
\label{sec:cl_res}
In this section, we compare \method{} with Lifelong nnU-Net, U-Net, and SOTA CL methods, including rehearsal-based approaches. However, it is important to note that rehearsal methods, while included in our comparison, pose significant impracticalities for medical CL settings. This is primarily due to stringent privacy policies concerning the storage of patient images and limited accessibility to historical data in continous setups.

\method{} achieves superior performances compared to the Lifelong nnU-Net and U-Net methods, by effectively leveraging domain-specific convolutional layers for accurate domain adaptation, Figure \ref{fig:spiders}. Table \ref{tab:ps} contains performances for all models and used CL methods.

\begin{figure*}[ht!]
    \centering
    \includegraphics[trim=0 5.5cm 9cm 0.35cm, clip, width=0.9\textwidth]{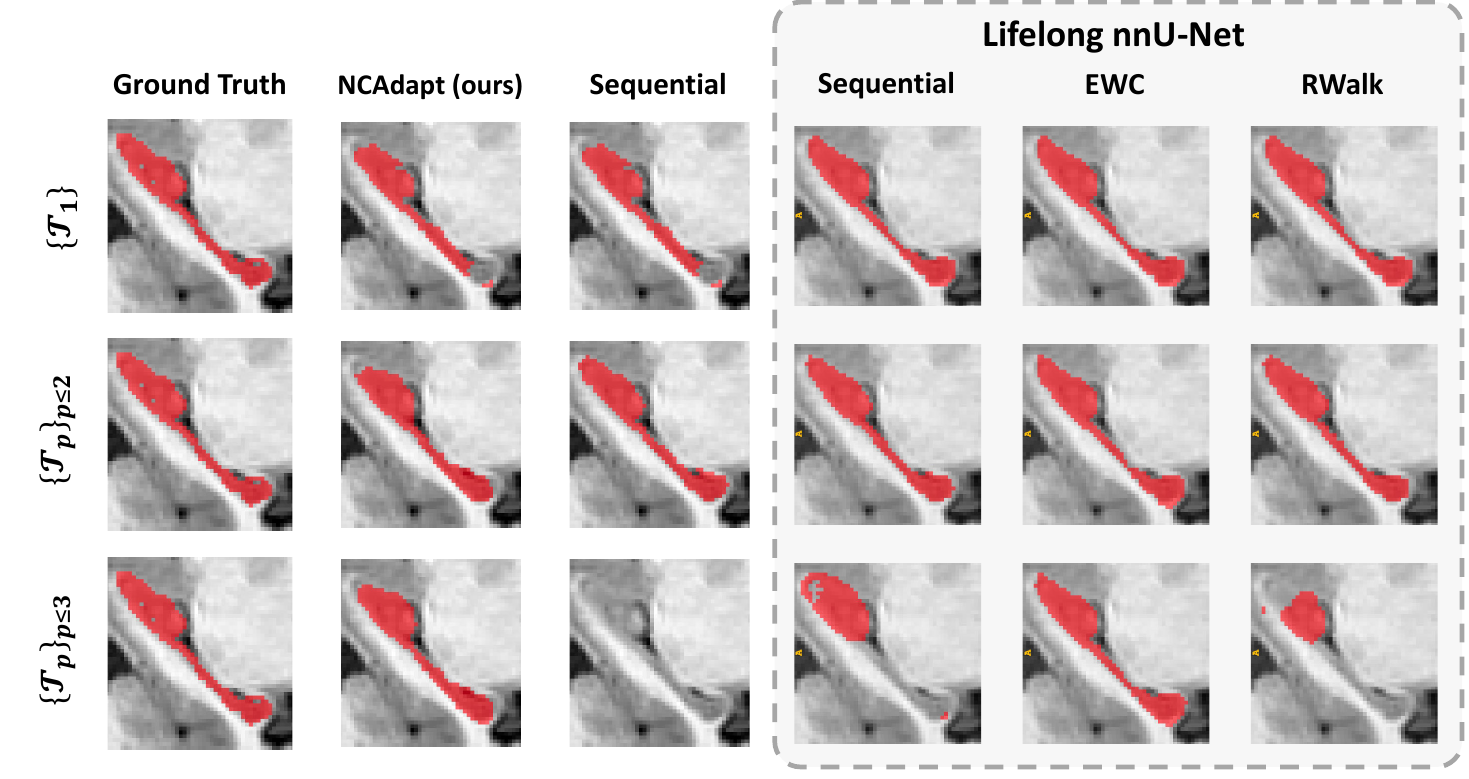}
    \caption{Temporal analysis for \method{}, sequential, EWC and RWalk using Case $s09\textunderscore{} L$ (3), Slice 27 from $\mathcal{T}_2$.}
    \label{fig:temp_res}
\end{figure*}

Lifelong nnU-Net realizes specific advantages depending on which CL method is used. However, it is notably surpassed by \method{}. When compared to the U-Net baselines, \method{} achieves comparable performance with substantially fewer parameters, while also effectively adjusting to varying image input sizes. This robust adaptation approach results in a notable performance increase of $46\%$ and $39\%$ in BWT and FWT, respectively, compared to the Lifelong nnU-Net with EWC. This contrasts with traditional methods that struggle to handle domain variations effectively over time.

wThese findings underscore the limitations of conventional methods, including advanced frameworks like Lifelong nnU-Net, which often fail to effectively manage domain variations over time for medical tasks. Compared to the traditional M3D-NCA in a sequential, EWC or even RWalk setup, \method{} achieves significantly better performances over all three metrices.

\subsection{Static model results}
\label{sec:base_models}
Table \ref{tab:baselines} shows that M3D-NCA's performance is comparable to U-Nets, showing competitive but not superior results compared to nnU-Net in individual, static task evaluations.  It shows baseline results for M3D-NCA, nnU-Net, and U-Net networks, each trained individually on each hippocampus tasks and evaluated across all tasks based on the Dice score.

\begin{table}[htp]
\begin{center}
\caption{Baseline results for M3D-NCA, nnU-Net and U-Net networks trained on every task individually and evaluated across all tasks; Bold values indicate the performance of the baseline on the validation set of the task it has been trained on.}
\label{tab:baselines}
\begin{adjustbox}{width=\linewidth}
{\begin{tabular}{ccccccc}
\toprule
\multicolumn{4}{c}{\multirow{2}{*}{Trained on}} & \multicolumn{3}{c}{Tested on -- Dice $\uparrow{ } \pm{ } $ $\sigma \downarrow $ {[}\%{]}} \\ \cmidrule{5-7}
& & & & HarP & Dryad & DecathHip \\ \midrule \midrule
\parbox[t]{2mm}{\multirow{3}{*}{\rotatebox[origin=c]{90}{\small{\shortstack{M3D-\\NCA}}}}} & & & HarP & $\mathbf{83.84 \pm 8.27}$ & $82.11 \pm 2.98$ & $7.61 \pm 8.64$ \\
& & & Dryad & $53.10 \pm 20.82$ & $\mathbf{89.05 \pm 1.17}$ & $0.00 \pm 0.00$ \\
& & & DecathHip & $2.41 \pm 5.12$ & $0.00 \pm 0.00$ & $\mathbf{83.79 \pm 4.30}$ \\ \cmidrule{1-7}
\parbox[t]{2mm}{\multirow{3}{*}{\rotatebox[origin=c]{90}{\small{nnU-Net}}}} & & & HarP & $\mathbf{88.81 \pm 9.06}$ & $87.70 \pm 0.95$ & $9.40 \pm 10.31$ \\
& & & Dryad & $33.23 \pm 30.28$ & $\mathbf{92.72 \pm 0.96}$ & $66.61 \pm 11.05$ \\
& & & DecathHip & $20.22 \pm 17.18$ & $56.65 \pm 5.68$ & $\mathbf{90.59 \pm 3.09}$ \\ \cmidrule{1-7}
\parbox[t]{2mm}{\multirow{3}{*}{\rotatebox[origin=c]{90}{\small{U-Net}}}} & & & HarP & $\mathbf{86.78 \pm 7.70}$ & $86.06 \pm 1.42$ & $31.55 \pm 20.19$ \\
& & & Dryad & $31.48 \pm 18.38$ & $\mathbf{50.51 \pm 12.5}$ & $31.48 \pm 18.38$ \\
& & & DecathHip & $43.88 \pm 26.68$ & $58.14 \pm 18.69$ & $\mathbf{94.91 \pm 2.86}$ \\ 
\bottomrule
\end{tabular}
}
\end{adjustbox}
\end{center}
\end{table}

While M3D-NCA performs well in static model scenarios, its true strength unfolds in continous setups with the ability to learn over time and adapt to new domains without a full re-training.

\subsection{\method{} ablation}
\label{sec:ablation}
In this ablation study, we systematically investigate various configurations of \method{} to assess the impact of freezing different layers within the M3D-NCA while preserving knowledge over time, shown in Table \ref{tab:ablation}:
\begin{itemize}
    \item \textit{$\text{NCA}_\text{FL}$} / \textit{$\text{NCA}_\text{HL}$} -- Freeze low-level or high-level NCA after initial task.
    \item \textit{$\text{NCA}_\text{FC}$}  -- Freezing M3D-NCA after initial task except perceiving convolution
    \item \textit{$\text{NCA}_\text{SA}$} -- Two convolutional layers added within every NCA (not domain-specific) that are only frozen during initial task.
\end{itemize}

\begin{table}[htb]
\centering
\caption{Ablation study comparing \method{} and its variants under different layer freezing configurations; best values are marked in bold.}
\label{tab:ablation}
\begin{adjustbox}{max width=\linewidth}
{\begin{tabular}{lccccc|cc}
\toprule
Method & Dice $\uparrow$ [\%] & BWT $\uparrow$ [\%] & FWT $\uparrow$ [\%] & \# Parameters (train) $\downarrow$\\ \midrule \midrule
$\text{nnU-Net}$ & $59.56 \pm 27.15$ & $-46.61 \pm 17.70$ & $\mathbf{-10.80 \pm 5.78}$ & 16,219,424 \\
\hline
$\text{NCA}_\text{FL}$ & $31.94 \pm 36.10$ & $-79.67 \pm 8.97$ & $-31.79 \pm 24.85$ & 8,768 \\
$\text{NCA}_\text{FH}$ & $56.41 \pm 18.66$ & $-42.87 \pm 9.79$ & $-44.75 \pm 37.81$ & \textbf{3,712} \\
$\text{NCA}_\text{FC}$ & $46.96 \pm 27.56$ & $-58.35 \pm 10.12$ & $-40.34 \pm 33.40$ & 5,952 \\
$\text{NCA}_\text{SA}$ & $29.10 \pm 40.21$ & $-87.28 \pm 3.17$ & $-13.66 \pm 2.80$ & 12,864 \\
$\text{\method{}}$ & $\mathbf{81.61 \pm 2.82}$ & $\mathbf{-0.43 \pm 0.10}$ & $-13.70 \pm 3.20$ & 6,336 \\
\bottomrule
\end{tabular}}
\end{adjustbox}
\end{table}

As shown in Table \ref{tab:ablation}, every method achieves different effects on the CL metrics, either increasing BWT, FWT or the mean Dice score. \method{} is the best-performing method across all ablations, by combining the benefits of each configuration. \method{} profits from all advantages, making it a robust solution for dynamic and evolving environments like in healthcare.

\subsection{Qualitative temporal evaluation}
\label{sec:qual}
To analyze the robustness of our proposed method, we illustrate segmentation masks in Figure \ref{fig:temp_res} for \method{} and sequential Lifelong nnU-Net with EWC and RWalk.

Additionally, \method{}'s lightweight nature and minimal parameter count significantly reduce computational requirements, making it a practical and efficient choice for CL applications in medical imaging. The model's ability to adapt seamlessly to varying domain characteristics ensures high performance across different imaging modalities and clinical settings.

\method{} consistently produces coherent segmentation masks across all training stages. In contrast, Sequential and RWalk for Lifelong nnU-Net lead to low-quality segmentations after training on the final stage 3 (${\mathcal{T}_{3}}$). This decline in performance on the sample scan for subsequent stages highlights the occurrence of catastrophic forgetting, wherein the network excessively adjusts its parameters to the most recent training data, exhibiting excessive plasticity. \method{} achieves robust predictions by maintaining quality across both early and later training stages, striking an \textbf{optimal balance with only 0.039\% of nnU-Net's parameters}.

This performance consistency of \method{} is attributed to its unique architecture that integrates domain-specific convolutional layers while maintaining a frozen NCA backbone. This design ensures that the general features learned during initial training stages remain intact, while the domain-specific layers adapt to new data without disrupting previously acquired knowledge. The careful balance between plasticity and stability in \method{} effectively mitigates the issues of catastrophic forgetting, a common challenge in lifelong learning scenarios.

\section{Conclusion}
\method{} introduces a novel approach utilizing NCA for CL in medical image segmentation, with a focus on hippocampus segmentation. This method effectively tackles domain adaptation challenges inherent in traditional U-Net architectures. Its superior adaptability and performance compared to the SOTA Lifelong nnU-Net framework, demonstrate its effectiveness in addressing these challenges with a fraction of the parameters used by nnU-Nets or U-Nets. By incorporating domain-specific convolutional layers and continuous adaptation, \method{} achieves robust segmentation performance across evolving datasets and domain shifts.

NCA-based methods, involve longer training times and significant VRAM during training due to the cell-based approach, which can be a limiting factor, particularly with high-resolution data. Nevertheless, the results highlight the potential of leveraging light-weight NCA-based approaches for CL in medical image segmentation. This approach offers a balanced solution between rigidity and plasticity without the need for additional CL methods that potentially decrease performance. The proposed method showcases significant improvements in handling dynamic and evolving medical imaging scenarios, ensuring reliable and accurate segmentation outcomes. 

This contribution not only advances the SOTA in medical image segmentation but also opens new avenues for developing efficient, adaptable, and resilient models for various clinical applications. The implications of this work extend beyond hippocampus segmentation, suggesting that NCA-based methods could be effectively applied to other challenging medical imaging tasks that require continual learning and domain adaptation.

\section{Reproducibility}
All datasets are publicly available and cited. Details on networks, data splits, and replication will be provided upon request. The code is publicly available under \url{https://github.com/MECLabTUDA/NCAdapt}.

{\small
\bibliographystyle{ieee_fullname}
\bibliography{main}

\begin{thebibliography}{10}\itemsep=-1pt

\bibitem{antonelli2021medical}
Michela Antonelli, Annika Reinke, Spyridon Bakas, Keyvan Farahani, Bennett~A Landman, Geert Litjens, Bjoern Menze, Olaf Ronneberger, Ronald~M Summers, Bram van Ginneken, et~al.
\newblock The medical segmentation decathlon.
\newblock {\em arXiv preprint arXiv:2106.05735}, 2021.

\bibitem{benjamin2018measuring}
Ari~S Benjamin, David Rolnick, and Konrad Kording.
\newblock Measuring and regularizing networks in function space.
\newblock {\em arXiv preprint arXiv:1805.08289}, 2018.

\bibitem{boccardi2015training}
Marina Boccardi, Martina Bocchetta, F{\'e}lix~C Morency, D~Louis Collins, Masami Nishikawa, Rossana Ganzola, Michel~J Grothe, Dominik Wolf, Alberto Redolfi, Michela Pievani, et~al.
\newblock Training labels for hippocampal segmentation based on the eadc-adni harmonized hippocampal protocol.
\newblock {\em Alzheimer's \& Dementia}, 11(2):175--183, 2015.

\bibitem{buzzega2020dark}
Pietro Buzzega, Matteo Boschini, Angelo Porrello, Davide Abati, and Simone Calderara.
\newblock Dark experience for general continual learning: a strong, simple baseline.
\newblock {\em Advances in neural information processing systems}, 33:15920--15930, 2020.

\bibitem{campello2021multi}
Victor~M Campello, Polyxeni Gkontra, Cristian Izquierdo, Carlos Martin-Isla, Alireza Sojoudi, Peter~M Full, Klaus Maier-Hein, Yao Zhang, Zhiqiang He, Jun Ma, et~al.
\newblock Multi-centre, multi-vendor and multi-disease cardiac segmentation: the m\&ms challenge.
\newblock {\em IEEE Transactions on Medical Imaging}, 40(12):3543--3554, 2021.

\bibitem{chaudhry2018riemannian}
Arslan Chaudhry, Puneet~K Dokania, Thalaiyasingam Ajanthan, and Philip~HS Torr.
\newblock Riemannian walk for incremental learning: Understanding forgetting and intransigence.
\newblock In {\em Proceedings of the European Conference on Computer Vision (ECCV)}, pages 532--547, 2018.

\bibitem{chaudhry2018efficient}
Arslan Chaudhry, Marc'Aurelio Ranzato, Marcus Rohrbach, and Mohamed Elhoseiny.
\newblock Efficient lifelong learning with a-gem.
\newblock {\em arXiv preprint arXiv:1812.00420}, 2018.

\bibitem{chen2021transunet}
Jieneng Chen, Yongyi Lu, Qihang Yu, Xiangde Luo, Ehsan Adeli, Yan Wang, Le Lu, Alan~L Yuille, and Yuyin Zhou.
\newblock Transunet: Transformers make strong encoders for medical image segmentation.
\newblock {\em arXiv preprint arXiv:2102.04306}, 2021.

\bibitem{codecarbon}
{CodeCarbon}.
\newblock Codecarbon: Measuring and mitigating the carbon footprint of your code.

\bibitem{de2021continual}
Matthias De~Lange, Rahaf Aljundi, Marc Masana, Sarah Parisot, Xu Jia, Ale{\v{s}} Leonardis, Gregory Slabaugh, and Tinne Tuytelaars.
\newblock A continual learning survey: Defying forgetting in classification tasks.
\newblock {\em IEEE transactions on pattern analysis and machine intelligence}, 44(7):3366--3385, 2021.

\bibitem{derakhshani2022lifelonger}
Mohammad~Mahdi Derakhshani, Ivona Najdenkoska, Tom van Sonsbeek, Xiantong Zhen, Dwarikanath Mahapatra, Marcel Worring, and Cees~GM Snoek.
\newblock Lifelonger: A benchmark for continual disease classification.
\newblock {\em -}, 2022.

\bibitem{diaz2018don}
Natalia D{\'\i}az-Rodr{\'\i}guez, Vincenzo Lomonaco, David Filliat, and Davide Maltoni.
\newblock Don't forget, there is more than forgetting: new metrics for continual learning.
\newblock In {\em Workshop on Continual Learning, NeurIPS 2018 (Neural Information Processing Systems}, 2018.

\bibitem{pmlr-v172-fuchs22a}
Moritz Fuchs, Camila Gonz\'{a}lez, and Anirban Mukhopadhyay.
\newblock Practical uncertainty quantification for brain tumor segmentation.
\newblock In Ender Konukoglu, Bjoern Menze, Archana Venkataraman, Christian Baumgartner, Qi Dou, and Shadi Albarqouni, editors, {\em Proceedings of The 5th International Conference on Medical Imaging with Deep Learning}, volume 172 of {\em Proceedings of Machine Learning Research}, pages 407--422. PMLR, 06--08 Jul 2022.

\bibitem{gilpin2019cellular}
William Gilpin.
\newblock Cellular automata as convolutional neural networks.
\newblock {\em Physical Review E}, 100(3):032402, 2019.

\bibitem{gonzalez2022task}
Camila Gonz{\'a}lez, Amin Ranem, Ahmed Othman, and Anirban Mukhopadhyay.
\newblock Task-agnostic continual hippocampus segmentation for smooth population shifts.
\newblock In {\em MICCAI Workshop on Domain Adaptation and Representation Transfer}, pages 108--118. Springer, 2022.

\bibitem{gonzalez2023lifelong}
Camila Gonz{\'a}lez, Amin Ranem, Daniel Pinto~dos Santos, Ahmed Othman, and Anirban Mukhopadhyay.
\newblock Lifelong nnu-net: a framework for standardized medical continual learning.
\newblock {\em Scientific Reports}, 13(1):9381, 2023.

\bibitem{gonzalez2020wrong}
Camila Gonzalez, Georgios Sakas, and Anirban Mukhopadhyay.
\newblock What is wrong with continual learning in medical image segmentation?
\newblock {\em arXiv preprint arXiv:2010.11008}, 2020.

\bibitem{hadsell2020embracing}
Raia Hadsell, Dushyant Rao, Andrei~A Rusu, and Razvan Pascanu.
\newblock Embracing change: Continual learning in deep neural networks.
\newblock {\em Trends in cognitive sciences}, 24(12):1028--1040, 2020.

\bibitem{isensee2021nnu}
Fabian Isensee, Paul~F Jaeger, Simon~AA Kohl, Jens Petersen, and Klaus~H Maier-Hein.
\newblock nnu-net: a self-configuring method for deep learning-based biomedical image segmentation.
\newblock {\em Nature methods}, 18(2):203--211, 2021.

\bibitem{jia2023importance}
Zhenge Jia, Jianxu Chen, Xiaowei Xu, John Kheir, Jingtong Hu, Han Xiao, Sui Peng, Xiaobo~Sharon Hu, Danny Chen, and Yiyu Shi.
\newblock The importance of resource awareness in artificial intelligence for healthcare.
\newblock {\em Nature Machine Intelligence}, 5(7):687--698, 2023.

\bibitem{joshi2012incremental}
Prachi Joshi and Parag Kulkarni.
\newblock Incremental learning: areas and methods-a survey.
\newblock {\em International Journal of Data Mining \& Knowledge Management Process}, 2(5):43, 2012.

\bibitem{kalkhof2023med}
John Kalkhof, Camila Gonz{\'a}lez, and Anirban Mukhopadhyay.
\newblock Med-nca: Robust and lightweight segmentation with neural cellular automata.
\newblock In {\em International Conference on Information Processing in Medical Imaging}, pages 705--716. Springer, 2023.

\bibitem{kalkhof2023m3d}
John Kalkhof and Anirban Mukhopadhyay.
\newblock M3d-nca: Robust 3d segmentation with built-in quality control.
\newblock In {\em International Conference on Medical Image Computing and Computer-Assisted Intervention}, pages 169--178. Springer, 2023.

\bibitem{kirkpatrick2017overcoming}
James Kirkpatrick, Razvan Pascanu, Neil Rabinowitz, Joel Veness, Guillaume Desjardins, Andrei~A Rusu, Kieran Milan, John Quan, Tiago Ramalho, Agnieszka Grabska-Barwinska, et~al.
\newblock Overcoming catastrophic forgetting in neural networks.
\newblock {\em Proceedings of the national academy of sciences}, 114(13):3521--3526, 2017.

\bibitem{kulaga2015multi}
Jessie Kulaga-Yoskovitz, Boris~C Bernhardt, Seok-Jun Hong, Tommaso Mansi, Kevin~E Liang, Andre~JW Van Der~Kouwe, Jonathan Smallwood, Andrea Bernasconi, and Neda Bernasconi.
\newblock Multi-contrast submillimetric 3 tesla hippocampal subfield segmentation protocol and dataset.
\newblock {\em Scientific Data}, 2(1):1--9, 2015.

\bibitem{mordvintsev2020growing}
Alexander Mordvintsev, Ettore Randazzo, Eyvind Niklasson, and Michael Levin.
\newblock Growing neural cellular automata.
\newblock {\em Distill}, 5(2):e23, 2020.

\bibitem{Pérez-García_2020}
Fernando Pérez-García.
\newblock fepegar/unet: Pytorch implementation of 2d and 3d u-net, 2020.

\bibitem{ranem2024continual}
Amin Ranem, Camila Gonz{\'a}lez, Daniel~Pinto dos Santos, Andreas~M Bucher, Ahmed~E Othman, and Anirban Mukhopadhyay.
\newblock Continual atlas-based segmentation of prostate mri.
\newblock In {\em Proceedings of the IEEE/CVF Winter Conference on Applications of Computer Vision}, pages 7563--7572, 2024.

\bibitem{ranem2022continual}
Amin Ranem, Camila Gonz{\'a}lez, and Anirban Mukhopadhyay.
\newblock Continual hippocampus segmentation with transformers.
\newblock In {\em Proceedings of the IEEE/CVF Conference on Computer Vision and Pattern Recognition}, pages 3711--3720, 2022.

\bibitem{sanner2021reliable}
Antoine Sanner, Camila Gonzalez, and Anirban Mukhopadhyay.
\newblock How reliable are out-of-distribution generalization methods for medical image segmentation?
\newblock In {\em DAGM German Conference on Pattern Recognition}, pages 604--617. Springer, 2021.

\bibitem{zenke2017continual}
Friedemann Zenke, Ben Poole, and Surya Ganguli.
\newblock Continual learning through synaptic intelligence.
\newblock In {\em International conference on machine learning}, pages 3987--3995. PMLR, 2017.

\end{thebibliography}
}

\clearpage

\setcounter{section}{0}
\setcounter{table}{0}
\setcounter{figure}{0}

\twocolumn[\section*{\centering Supplementary Material}]

\section{Cardiac Datasets}
We focus on the continual segmentation of the left ventricle (LV) in cardiac MRIs, utilizing data from the Multi-Centre, Multi-Vendor \& Multi-Disease Cardiac Image Segmentation Challenge (M\&Ms) \cite{campello2021multi}, Table \ref{tab:data}. This dataset includes 75 labeled cases acquired with Siemens scanners and 75 cases acquired with Philips scanners. Although the original dataset includes annotations for the left and right ventricles as well as the myocardium, our study specifically targets LV segmentation.

\begin{table}[htp]
\centering
\caption{Image and label characteristics of the used cardiac datasets.}
\label{tab:data}
\begin{adjustbox}{width=\linewidth}{
\begin{tabular}{ccccccc}
\hline
\multicolumn{1}{l}{Dataset} & & Siemens           & &  Philips \\ \hline \hline
\multicolumn{1}{l|}{\# Cases}  & & 75              & & 75    \\
\multicolumn{1}{l|}{Resolution} & & {[}12 256 256{]}    & & {[}12 256 256{]}  \\
\bottomrule
\end{tabular}
}
\end{adjustbox}
\end{table}

The multi-class nature of the original problem, which involves different anatomical structures, allows us to investigate how segmentation performance varies depending on the shape and size of the region of interest. However, by focusing solely on the LV, we concentrate on the challenges posed by this particular structure while ensuring that the method is robust to variations across different scanners and patient populations. For each dataset, we maintain a consistent split by using 20\% of the data for testing purposes across all experiments.\\
This targeted approach allows us to effectively evaluate the adaptability and performance of our method in a controlled yet challenging scenario, demonstrating the potential benefits of continual learning (CL) in hippocampus and cardiac image segmentation.

\section{CL performance for LV segmentation}

In this section, we evaluate the CL performance of our method specifically for left ventricle (LV) segmentation in cardiac MRIs. The emphasis is on evaluating how effectively the model generalizes and adapts to new tasks while preserving knowledge from previous tasks beyond the initial hippocampus segmentation task from the main manuscript. We utilize backward transfer (BWT) and forward transfer (FWT) metrics defined in the main manuscript to measure the model's ability to learn incrementally and avoid catastrophic forgetting.

\begin{figure}[htb!]
    \centering
    \includegraphics[width=0.475\textwidth]{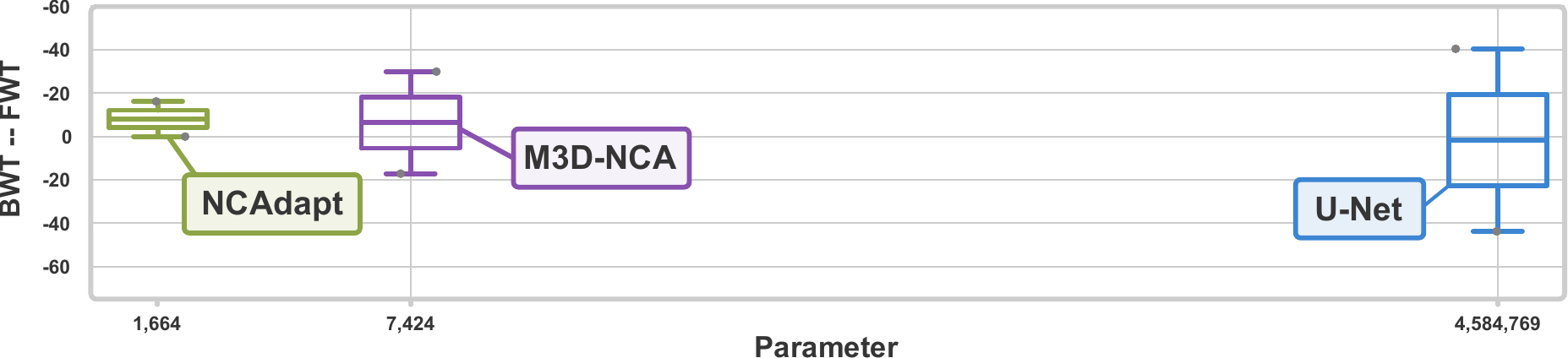}
    \caption{Comparison of BWT and FWT performance against number of parameters for \method{} and sequential U-Net trained on the Cardiac datasets; smaller boxes indicate superior performance. Note: Values above zero on the y-axis represents FWT and values below BWT.}
    \label{fig:bwt_fwt}
\end{figure}

To provide a comprehensive evaluation of continual learning (CL) performance, Figure \ref{fig:bwt_fwt} presents a comparison of backward transfer (BWT) and forward transfer (FWT) metrics across different models, including \method{}, sequential U-Net, and TransU-Net, all trained on the Cardiac datasets. The figure highlights the relationship between the number of parameters and the CL performance, with smaller boxes indicating superior performance. This visual comparison helps illustrate how each model balances the trade-off between model complexity and learning efficiency. Notably, positive values on the y-axis represent FWT, while negative values represent BWT, offering insights into how well the models generalize to new tasks while retaining knowledge from previous ones.\\
In addition, Table \ref{tab:ps} summarizes the overall CL performance of the final models on the Cardiac datasets, including mean Dice scores, BWT and FWT metrics, along with the total number of trainable parameters, training runtime, and inference time. Since the setup involves two datasets, only a single BWT and FWT value for each model can be reported, i.e. with standard deviation \(0\%\). This table provides a detailed comparison of the models, with the best values for the sequential setup highlighted in bold. Methods marked with an asterisk (*) are rehearsal-based, indicating that they leverage previously seen data during training to improve CL performance. This table serves as a key reference for understanding the trade-offs between performance, computational cost, and memory efficiency in the context of LV segmentation.

\begin{table*}[ht!]
\caption{CL performance of the final model on the Cardiac datasets; mean Dice, BWT and FWT over all tasks including standard deviation, total amount of trainable parameters, training runtime and inference time in seconds; best values for sequential setup are marked in bold. Methods marked with * are rehearsal-based methods.}
\label{tab:ps}
\begin{adjustbox}{max width=\linewidth}
{\begin{tabular}{lccccc|ccc}
\toprule
Method & Fixed param & Tuned param & Dice $\uparrow$ [\%] & BWT $\uparrow$ [\%] & FWT $\uparrow$ [\%] & \# Parameters (train) $\downarrow$ & Runtime [sec] $\downarrow$ & GPU sec $\downarrow$ \\ \midrule \midrule
$\text{Sequential}_{\text{U-Net}}$ & \multirow{3}{*}{--} & \multirow{3}{*}{--} & $57.63 \pm 33.24$ & $-43.73$ & $48.97$ & 4,584,769 & $\mathbf{8.63}$ & $1.05$ \\
$\text{Sequential}_{\text{NCA}}$ & & & $72.67 \pm 15.76$ & $-17.10$ & $-29.98$ & 7,424 & $24.0$ & $1.20$ \\
$\text{Sequential}_{\text{\method{}}}$ & & & $71.85 \pm 1.83$ & $\mathbf{-0.001}$ & $-16.22$ & \textbf{1,664} & $19.16$ & $\mathbf{0.123}$ \\
\midrule
$\text{EWC}_{\text{NCA}}$ & -- & $\lambda = 0.4$ & $71.28 \pm 17.29$ & $-18.60$ & $-26.24$ & 7,424 & $23.08$ & $1.15$ \\
$\text{RWalk}_{\text{NCA}}$ & $\alpha = 0.9$ & $\lambda = 0.4$ & $67.58 \pm 6.15$ & $0.919$ & $-23.90$ & 7,424 & $23.59$ & $1.30$ \\
\midrule
$\text{SI}^{*}_{\text{U-Net}}$ & $c = 0.4$ & \multirow{4}{*}{--} & $56.30 \pm 33.42$ & $-45.13$ & $-39.07$ & \multirow{4}{*}{4,584,769} & $8.80$ & $1.11$\\
$\text{FDR}^{*}_{\text{U-Net}}$ & -- & & $58.29 \pm 31.75$ & $-47.24$ & $-45.28$ & & $9.54$ & $1.14$ \\
$\text{DER}^{*}_{\text{U-Net}}$ & $\alpha = 0.4$ & & $59.70 \pm 28.94$ & $-33.15$ & $-39.33$ & & $10.04$ & $1.00$ \\
$\text{AGem}^{*}_{\text{U-Net}}$ & -- & & $59.33 \pm 29.71$ & $-33.70$ & $-49.41$ & & $10.15$ & $1.00$ \\
\bottomrule
\end{tabular}}
\end{adjustbox}
\end{table*}

\end{document}